\documentclass[10pt, conference, letterpaper]{IEEEtran}

\usepackage[cmex10]{amsmath}
\usepackage{amssymb}
\usepackage{cite}
\usepackage{graphicx}
\usepackage{epsfig}
\usepackage{url}
\usepackage{color}
\usepackage{setspace,epstopdf,subfigure}
\usepackage{verbatim}
\usepackage{multirow}
\usepackage{array}
\usepackage{float}
\usepackage{amsthm}
\usepackage{listings}
\usepackage{xcolor}
\usepackage{algorithm}
\usepackage{algpseudocode}
\usepackage{pifont}
\usepackage{caption}
\usepackage{booktabs}
\usepackage{threeparttable}
\usepackage[square, comma, sort&compress, numbers]{natbib}
\usepackage{hyperref}

\definecolor{dkgreen}{rgb}{0,0.6,0}
\definecolor{gray}{rgb}{0.5,0.5,0.5}
\definecolor{mauve}{rgb}{0.58,0,0.82}

\newcommand{\ignore}[1]{}

\newcommand{\red}[1]{{\color{red} #1}}

\theoremstyle{definition}

\newtheorem*{defn*}{Definition}
\newtheorem*{atoms*}{Atoms}
\newtheorem*{rules*}{Rules}
\newtheorem*{bfeatures*}{Binary Features}
\newtheorem*{qfeatures*}{Quantative Features}
\newtheorem*{theorem*}{Theorem}

\newcommand\MyBox[2]{
  \fbox{\lower0.75cm
    \vbox to 1.7cm{\vfil
      \hbox to 1.7cm{\hfil\parbox{1.4cm}{#1\\#2}\hfil}
      \vfil}%
  }%
}
\renewcommand{\baselinestretch}{.98}
\usepackage[belowskip=-1pt,aboveskip=5pt]{caption}

\begin{document}

\title{LeakSemantic: Identifying Abnormal Sensitive Network Transmissions in Mobile Applications}

\author{
    \IEEEauthorblockN{
    Hao Fu\IEEEauthorrefmark{1}, 
    Zizhan Zheng\IEEEauthorrefmark{2}, 
    Somdutta Bose\IEEEauthorrefmark{1}, 
    Matt Bishop\IEEEauthorrefmark{1}, 
    Prasant Mohapatra\IEEEauthorrefmark{1}}
    \IEEEauthorblockA{\IEEEauthorrefmark{1}Department of Computer Science, University of California, Davis, USA.}\vspace{-3.5mm} \\
    \IEEEauthorblockA{\IEEEauthorrefmark{2}Department of Computer Science, Tulane University, New Orleans, USA.}\vspace{-3.5mm} \\
    \texttt{\{haofu, sombose, bishop, pmohapatra\}@ucdavis.edu, zzheng3@tulane.edu}
}

\maketitle

\begin{abstract}
  Mobile applications (apps) often transmit sensitive data through network with various intentions.
  Some transmissions are needed to fulfill the app's functionalities.
  However, transmissions with malicious receivers may lead to privacy leakage and tend to behave stealthily to evade detection.
  The problem is twofold: how does one unveil sensitive transmissions in mobile apps, and given a sensitive transmission, how does one determine if it is legitimate?

  In this paper, we propose LeakSemantic, a framework that can automatically locate abnormal sensitive network transmissions from mobile apps.
  LeakSemantic consists of a hybrid program analysis component and a machine learning component.
  Our program analysis component combines static analysis and dynamic analysis to precisely identify sensitive transmissions. 
  Compared to existing taint analysis approaches, LeakSemantic achieves better accuracy with fewer false positives and is able to collect runtime data such as network traffic for each transmission.
  Based on features derived from the runtime data, machine learning classifiers are built to further differentiate between the legal and illegal disclosures.
  Experiments show that LeakSemantic achieves 91\% accuracy on 2279 sensitive connections from 1404 apps.
\end{abstract}

\section{Introduction} \label{sec:intro}
The exponential growth of mobile devices has raised significant security concerns.
Due to the large amount of sensitive data saved on these devices and the coarse-grained permission management in mobile systems,
they are vulnerable to various privacy and malicious infringing behaviors, which is often hard to detect by mobile users themselves.
One reason is that malicious apps have begun taking steps to avoid detection by introducing {\it logic bombs}~\cite{yangappcontext}.
For instance, an app can hide malicious transmissions by receiving certain commands from remote servers.
Even if a sensitive network transmission is known, an end user often has trouble telling if it is necessary since the legitimacy of a sensitive transmission depends on its purpose.
\ignore{\red{Fig.~\ref{fig_example} description here?}}
Therefore, it is critical to uncover security-sensitive behaviors and understand the {\it intention} of them to detect the abnormal ones.

\ignore{
\begin{figure}
\centering
\includegraphics[height=3.0in,width=0.32\textwidth]{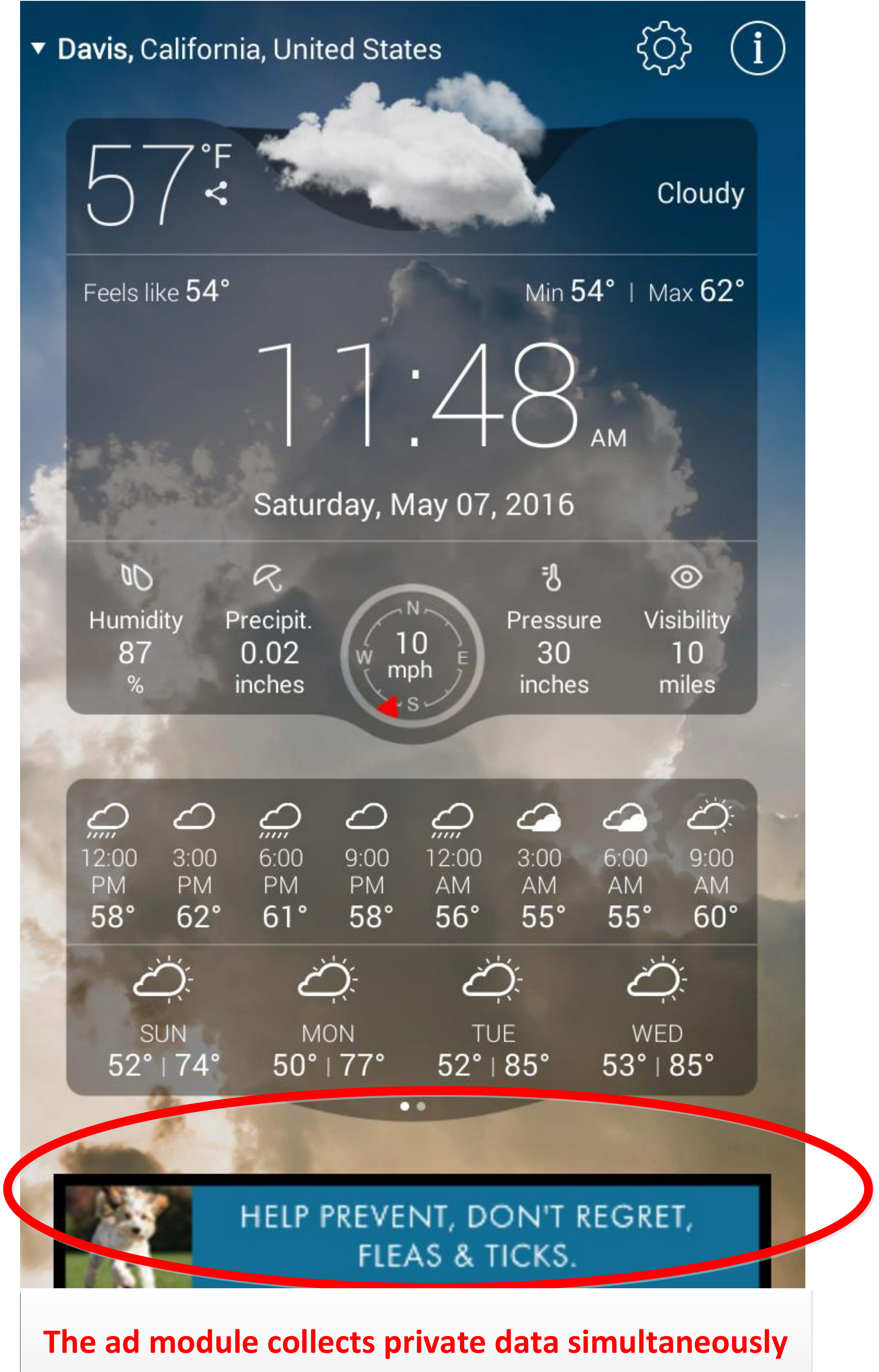}
\caption{An example}
\label{fig_example}
\end{figure}
}
In this paper, we focus on detecting abnormal sensitive network transmissions in Android apps.
These transmissions either leak user private data to malicious servers, or collect sensitive information for purposes such as advertisements that do not contribute to fulfill the functionalities of the underlying apps.
Despite the fast-growing literature on mobile device security and privacy, existing approaches are insufficient for identifying abnormal sensitive network behaviors.
In particular, their ability is limited by the complexity of the Android API and runtime, which involves millions of lines of code.
Moreover, they focus on detecting sensitive transmissions only and are often not able to distinguish between normal and abnormal sensitive transmissions.
\ignore{
Moreover, they mainly rely on static program analysis, which incurs high overhead and is often not applicable to large apps.
On the other hand, existing network traffic based approaches mainly build their classifier from traffic data collected {\it passively}.
The samples thus collected cannot provide a good coverage of network behavior.
Moreover, the mixture of insensitive traffic and sensitive traffic in the samples increases the learning burden and decreases the accuracy.
}

To address these limitations, we propose {\bf LeakSemantic}, a novel approach that combines program analysis and machine learning to identify abnormal sensitive network transmissions more accurately through a better understanding of network semantics.
LeakSemantic adopts a hybrid static-dynamic analysis approach to uncover sensitive transmissions (both normal and abnormal).
The hybrid approach not only produces better results than a purely static or dynamic analysis approach, but is also able to generate network traffic data in a proactive way, which provides a better characterization of network behavior than widely-used static program analysis based approaches~\cite{choi2015extractocol, huang2014asdroid, lu2015checking, chen2015droidjust}. 
For example, the hostname of the malicious server in the {\tt PJAPPS} malware family is encrypted as ``\texttt{ax3mkl4mgele2guoo9f1hc3ohm}" and the real address (\texttt{xml.meego91.com}) is only revealed at runtime. 
Without running the code, static analysis methods fail to decrypt the malicious hostname, which is an important feature for detecting abnormal network transmissions. 
Instead, the dynamic execution used by LeakSemantic enables tracking runtime information including decrypted hostnames.
The traffic data generated are then fed to the machine learning component to build classifiers for detecting abnormal transmissions.
Note that our program analysis component can potentially gather more features than just network traffic, which can be useful to differentiate between normal and abnormal flows.
We focus on network traffic in this work so that the learning model thus built can be applied even when the app code is not available, e.g., when it is integrated into a network-based intrusion detection system. 
Thus, LeakSemantic can be used in various settings.  
When deployed locally, it allows app market operators to identify privacy leakage in an app before releasing it to the market.
Moreover, network administrators can benefit from the detection model constructed by LeakSemantic to protect users from unintended transmissions.  

A major challenge of program analysis for mobile apps is how to achieve both accuracy and precision. 
Static program analysis examines the program dependencies in mobile apps without actually executing them. 
Because of its static nature, it cannot handle reflective calls whose target class or method name is concatenated at runtime, and loading code dynamically is becoming more common~\cite{lindorfer2014andrubis}.
Static analysis also introduces false alarms as an over-estimated method.
In contrast, dynamic analysis chases the runtime behavior of apps and is applicable even when reflection is present.
Unlike static analysis that explores all code paths including infeasible ones, 
dynamic analysis only proceeds to feasible paths and therefore introduces lower false positive rate.
Moreover, it can obtain data that are not available in the static setting, such as network traffic data using encrypted URLs.
However, by focusing only on the runtime behaviors, dynamic analysis suffers from insufficient coverage and hence false negatives.

Recent research efforts aim to combine static and dynamic program analysis to ameliorate the above problems~\cite{appaudit, rasthofer2016harvesting, wong2016intellidroid}.
We continue this line of research and propose a novel design of hybrid program analysis.
LeakSemantic adopts light-weight static analysis to flag potential vulnerabilities, and creates an environment to dynamically confirm the suspicions.
Our static analysis provides precise modeling of the call relationships inside an app component, which is crucial for the integrated dynamic analysis component.
We introduce a new execution trace generation technique that enables LeakSemantic to uncover malicious behaviors on which previous studies would fail. As we will show in Section~\ref{sec:dynamic}, it is insufficient to simply identify code paths leading to targeted APIs.
To this end, LeakSemantic dynamically spreads the code coverage and computes the appropriate traces to trigger stealthy behaviors.
It also takes into account various sources of unknown variables with an effective handling of unknowns, which further reduces the number of false negatives.

\ignore{
We have applied our approach to detect abnormal sensitive network transmissions, that is, sensitive communications without clear purpose such as private data leakage.
Although our approach currently focuses on network resources, the main idea can be generalized to detect the abnormal use of other types of resources.
}

To summarize, this paper presents the following contributions:
\begin{itemize}
\item We propose a novel hybrid static-dynamic program analysis technique to locate sensitive network transmissions in mobile applications.
Our approach not only enables better accuracy\ignore{efficiency-accuracy tradeoff} and precision, but also helps derive more detailed features, e.g., traffic URLs, that are important for network behavior analysis.
\item We present the design and implementation of LeakSemantic, a detection system that combines program analysis and machine learning to identify networking related abnormal behavioral patterns.
 Instead of classifying a whole app as malicious or not as most previous work does, our approach is able to distinguish malicious behavior from normal behavior within an app.
 We also show that network-level detection can benefit from the information provided by program analysis.
\item We evaluate the effectiveness of LeakSemantic using two micro-benchmark suites and 1404 real-world apps.
Our hybrid program analysis produces better results than any of the three state-of-the-art taint analysis tools used in evaluations.
Experiments further show that LeakSemantic is fast and cheap, allowing it to identify true threats inside the real apps with high accuracy.
\end{itemize}

The rest of the paper is organized as follows. We highlight our system overview in Section~\ref{sec:overview}.
The technical details are included in Section~\ref{sec:sys}.
After presenting the system implementation in Section~\ref{sec:implementation}, we show the evaluation results in Section~\ref{sec:eval}.
We discuss the limitations in Section~\ref{sec:limitation} and the related work in Section~\ref{sec:related}.
Finally, Section \ref{sec:conclusion} concludes the paper.

\begin{figure}[htb]
\centering
\includegraphics[height=1.5in,width=0.38\textwidth]{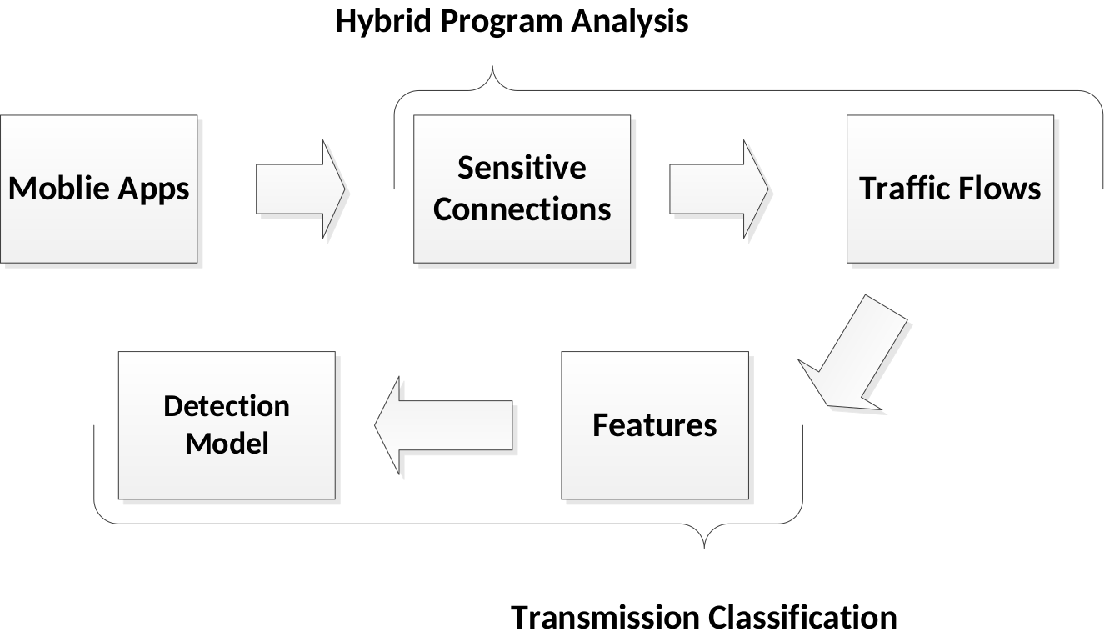}
\caption{\small System Architecture}
\label{fig_overview}
\end{figure}

\section{Overview} \label{sec:overview}
Figure~\ref{fig_overview} depicts the architecture of LeakSemantic.
From the datasets of authentic apps and malwares, our system proceeds in the following steps:
\begin{itemize}
  \item{\bf Hybrid Program Analysis:}
  The phase of hybrid program analysis precisely identifies and characterizes the leaking connections in the target app.
  We first perform static analysis to retrieve the call graphs of the corresponding app.
  To better model the lifecycles of app components and runtime events, we create {\tt dummyMain()} for each component.
  The invocations of sensitive APIs (sources) that collect private data with their entry points are identified through traversal of the graphs.
  We then construct execution traces and run the program from the set of traces.
  The information flow analysis is performed during the execution.
  If a connection point (sink) is reached, we record the dynamic data of the communication.
  To achieve better coverage, we have designed methods to generate execution traces and handle unknowns encountered during runtime.
  \item {\bf Transmission Classification:}
  Having extracted traffic information about the sensitive connections, we then derive a set of features that can be used by the anomaly detection system.
  Concretely, we concentrate on building machine learning classifiers using lexical features derived from URLs. Our novel design enables us to build models for both host-based and network-based detection.
\end{itemize}

\section{LeakSemantic} \label{sec:sys}
To model the runtime behavior of apps while achieving good coverage, we use a hybrid program analysis that combines static analysis and dynamic analysis.
In Android, a medium-sized app can contain dozens of components and thousands of methods. Dynamic traversal of all possible paths is expensive and infeasible in practice.
Our approach leverages light-weight static analysis to locate invocations of sensitive APIs and the corresponding components.
The output of static analysis will help guide dynamic analysis.
Machine learning models are then constructed with the flows derived by dynamic analysis.
It is crucial that LeakSemantic can generate sensitive flows with decrypted URLs.
Finding 1 in Section~\ref{sec:eval} states that the detection ratio decreases obviously if the training data does not cover sufficient characteristics of the malicious flows.
\begin{figure}[htb]
\centering
\includegraphics[height=1.6in,width=0.2\textwidth]{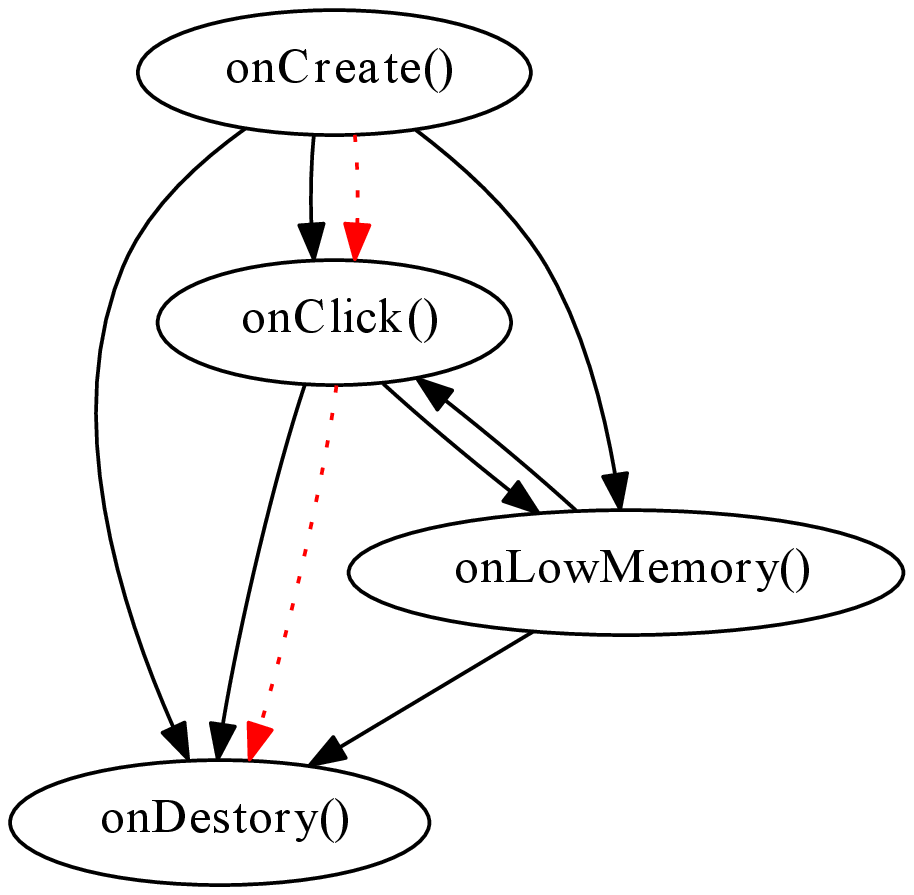}
\vspace{-.5ex}
\caption{\small
The call graph of \texttt{Activity1} modeled by the corresponding \texttt{dummyMain()}.
The solid lines indicate call relationships among the callbacks and the dashed lines specify one possible execution trace on the call graph.}
\label{fig_cfg}
\end{figure}

\subsection{Static Analysis} \label{sec:static}
\vspace{-1ex}
Static analysis is responsible for constructing the call graph of the target app, which guides the upcoming dynamic analysis.
Unlike (desktop) programs written in \texttt{C} that contain a main function as the entry point of the execution,
Android applications do not contain a single main method.
Instead, they are composed of multiple components, where each \texttt{Activity} or \texttt{Service} component is a \texttt{Java} class and has its own lifecycle and event listeners.
The lifecycle models transitions such as creation, pause, resume, and termination, between the states of a component.
Event listeners allow applications to respond to various types of runtime events such as UI interactions or receiving SMS.
The lifecycle and event listeners are constructed from the corresponding callback methods and every callback can be treated as an entry point because they are implicitly called by the Android framework.

To construct call graphs of applications, previous work typically creates one or more dummy main routines that are shared by multiple components.
For example, FlowDroid \cite{arzt2014flowdroid} creates a single dummy main for the entire application and all components share that main.
AppAudit~\cite{appaudit} introduces a shared dummy main for all components of the same category (\texttt{Activity} or \texttt{Service}).
However, analyses starting from a shared dummy main may include components that do not contribute to leakage.
Moreover, a shared dummy main blurs the connections between event listeners and components.
It is possible that an event listener may be linked to the wrong component so that the latter can directly invoke the former during the analysis, even though this would not happen in a real setting. 
Instead of constructing a shared dummy main, we let each component have its own dummy main to eliminate the confusion and alleviate the overhead of dynamic analysis.
Each component thus has a call graph (an example is given in the Figure~\ref{fig_cfg}).
The event listeners such as {\tt onClick()} and {\tt onLowMemory()} embedded with the component are registered after {\tt onCreate()}.
\texttt{onClick()} is a UI callback that is invoked once the corresponding buttons are clicked, whereas \texttt{onLowMemory()} is called once the available memory of the device is lower than a threshold value.

A \textbf{source} is an invocation of an API provided by the Android framework to retrieve the sensitive information from the underlying device.
We use the list from Susi~\cite{rasthofer2014machine} to locate the sources.
An example source is the invocation of \texttt{getDeviceID()} at line 13 shown in Listing~\ref{lst:component}.
The program is inspired by {\tt EventOrdering1} in DroidBench~\cite{arzt2014flowdroid}.
For each source, the corresponding entry point of the component in the call graph is extracted with applying a graph traversal algorithm on the call graph.
For instance, the entry point {\tt onClick()} of the component \texttt{Activity1} in Listing~\ref{lst:component} is located through breadth-first search beginning with \texttt{getDeviceID()} on the call graph.
The entries with relevant call graphs serve as the starting points of dynamic analysis.
We will explain this in detail in the next subsection.

\begin{lstlisting}[caption=An example component, label=lst:component, float, floatplacement=H]
  class Acvitity1 extends Activity {
        String url = ""
        String imei = "";
        String tmp = "";

        void onCreate() {
          /* initiate the activity */
          ...
          url = "gongfu188.com";
        }

        void onClick() {
          tmp = <get phone manager>.getDeviceId(); // source
        }

        void onLowMemory() {
          url = url.concat(imei);
          URLConnection conn = new URL(url).openConnection(); // sink
          imei = tmp; // tainted
        }

        void onDestroy() {
          /* finish the activity */
          ...
        }
  }
\end{lstlisting}

\ignore{
\red{Reduce unknowns, and performance issue}

\red{Multi-threading?}

\red{One challenge of implementing static analysis in Android is to handle Inter-Component Communication (ICC). In our hybrid analysis framework, we use static analysis to extract call graphs for all the components and defer ICC mapping to the dynamic analysis (see Section~\ref{} for details).} Once we obtain the call graph, a graph traversal algorithm (e.g., breadth-first search) is applied to locate the set of sources, i.e., the invocations of sensitive APIs. For each sensitive invocation, the entry point of the corresponding component where the invocation lies is also extracted. These entry points serve as the starting points of the dynamic analysis.
We will explain the detail in the next subsection.
}

\subsection{Dynamic Analysis} \label{sec:dynamic}
The dynamic analysis component of LeakSemantic consists of an executor with a taint analysis module and a simulation of the Android runtime.
The executor is our own version of the Dalvik virtual machine.
It is able to directly unpack Android package files and execute the bytecode instructions.
We feed a set of traces to the executor.
The execution traces are derived not only from the results of the static analysis, but also from the execution procedure itself.
The novel design enables capturing the misbehavior missed by state-of-the-art approaches, which we will discuss in detail later.
During the execution, whenever a sensitive source API is invoked, the taint analysis module starts to track the propagation of sensitive values associated with the source API.
When one or more sensitive values reach a network connection API call (a \textbf{sink}) such as {\tt URL/openConnection()} in line 18 in Listing~\ref{lst:component}, which implies that the transmission is \textbf{sensitive}, the corresponding runtime information such as the network traffic data is recorded.
We adopt general taint policies used in previous work~\cite{enck2014taintdroid, appaudit} to specify the propagation procedure.
For example, one rule set $x$ is tainted as long as one of the operands in the instruction ``$x = y\text{ binop }z$" is tainted.
To improve the accuracy of the data flow analysis, 
we have further developed libraries to emulate the fundamental behaviors of the Android runtime.
The implementation details are described in Section~\ref{sec:implementation}.
In the following, we discuss how LeakSemantic constructs execution traces and how it handles unknown values during the analysis. 

\subsubsection{Execution Trace Generation}
We leverage the outcomes of the static analysis phase to derive a set of \textit{basic} execution traces, where each trace is a sequence of specific API calls beginning with a lifecycle callback and ending with an APT call where a source is triggered. 
For instance, for the entry point \texttt{onClick()} in \texttt{Activity1}, LeakSemantic builds an execution trace \texttt{onCreate()} $\rightarrow$ \texttt{onClick()} that informs the executor to invoke \texttt{onClick()} after calling \texttt{onCreate()}.
The execution trace is generated by applying depth-first search to find a path from \texttt{onCreate()} to \texttt{onClick()} in the call graph (Figure~\ref{fig_cfg}).
The default values of global variables are normally initialized at the lifecycle callbacks such as \texttt{onCreate()} and \texttt{onStart()}.
We choose to execute from these callbacks to reduce unknown variables, which in turn reduces unknown branches that need to be explored and improves the efficiency of dynamic analysis. 
Properly modeling the unknowns is challenging in general and will be discussed in more detail in the following subsections.
In addition to reducing unknowns, 
our approach also enables LeakSemantic to generate more complete URLs, which is important for building accurate classifiers (see Section~\ref{sec:classification}). 
As we can see in Listing~\ref{lst:component}, the connection in line 18 can only be correctly triggered if {\tt url} is properly assigned with the hostname in line 9.

The \textit{de facto} hybrid analysis approaches such as AppAudit, Harvester~\cite{rasthofer2016harvesting} and IntelliDroid~\cite{wong2016intellidroid} only use code paths with certain code locations (e.g., a sink) and terminate the analysis once one such location is reached.
However, reachability alone does not necessarily imply the exposure of true malicious behavior.
Reconsider the code snippet shown in Listing~\ref{lst:component}.
A direct invocation of \texttt{onLowMemory()} does not lead to a leakage since the argument of the sink in line 18 may have an empty \texttt{imei}. 
Given that \texttt{tmp} is tainted in \texttt{onClick()}, the correct order to trigger a real leakage is to invoke \texttt{onLowMemory()} twice. The corresponding execution sequence can be represented as \texttt{onCreate()} $\rightarrow$ \texttt{onClick()} $\rightarrow$ \texttt{onLowMemory()} $\rightarrow$ \texttt{onLowMemory()}. 

To correctly generate the set of execution traces that trigger the actual leakages (or other types of abnormal behavior), we parse the code of the executable callbacks to determine whether they contain statements that read the corresponding newly tainted variables.
A new execution trace is then constructed by expanding the existing trace with relevant callbacks.
For instance, after executing the trace \texttt{onCreate()} $\rightarrow$ \texttt{onClick()}, \texttt{onLowMemory()} is identified since it reads the value from the tainted variable \texttt{tmp}.
A new execution trace \texttt{onCreate()} $\rightarrow$ \texttt{onClick()} $\rightarrow$ \texttt{onLowMemory()} is created.
Similarly, LeakSemantics constructs \texttt{onCreate()} $\rightarrow$ \texttt{onClick()} $\rightarrow$ \texttt{onLowMemory()} $\rightarrow$ \texttt{onLowMemory()} once finishing running \texttt{onCreate()} $\rightarrow$ \texttt{onClick()} $\rightarrow$ \texttt{onLowMemory()}.
We can set a threshold on the number of execution traces to save analysis time in practice.

\begin{lstlisting} [caption=A logic bomb, label=lst:dfs, float, floatplacement=H]
  String mRun = getSearchTask(); // commands
  void doSearchTask() {
      if (mRun == null) {
          reportState(1);
          if (mRun != null) {
              runPackage(mPkgName); // leak
          } else {
            ...
          }
      } else {
          ...
      }
}
\end{lstlisting}

\subsubsection{Sources of Unknowns} \label{sec:unknowns}
During the execution, the dynamic analysis may encounter unknown variables that have no explicit assigned value to the executor.
As mentioned earlier, running from \texttt{onCreate()} alleviates the issue through initializing the component as completely as possible.

In addition to the above mentioned unknowns, we observe that there are many cases where the accurate value of a variable is dependent on the runtime context, which can be categorized as follows:
\begin{itemize}
  \item User input: input from end users during the interactions with the user interface;  
  \item Device status: the real time status, such as WiFi on/off and the power level, of the underlying device; 
  \item Natural environment: e.g., current temperature, coordinate and time; 
  \item Incoming information: the content of the SMS and the network responses received while using the app. 
\end{itemize}
Malicious apps may hide their behavior by leveraging some of the factors mentioned above to create malicious code that is only triggered under certain circumstances.
For instance, \texttt{RCSAndroid} waits for incoming SMS messages and checks whether these messages contain specific commands and then decides whether to transmit the user data~\cite{fratantonio2016triggerscope}, and the \texttt{DroidDream} malware family 
triggers its malicious payload only at night \cite{yangappcontext}.
As another example, consider the code shown in Listing~\ref{lst:dfs}, which comes from a malware sample of the \texttt{DroidKunfu1} family.
In line 1 the program contacts a remote control server and retrieves the commands into {\tt mRun}.
\texttt{reportState()} is responsible for collecting user private data and it is only triggered when the malicious server replies with certain characters. 
In other words, the dynamic context causes the executor to generate different outcomes even for the same input trace.
To detect such malicious behavior, it is therefore important to treat those variables whose values vary over the context as unknowns. 

\subsubsection{Handling of Unknowns}
To represent the set of variables with unknown values, we maintain a symbolic state $\sigma$ that maps variables to symbolic expressions, and a symbolic path constraint $PC$,
which is a quantifier-free first-order formula over symbolic expressions.
Both $\sigma$ and $PC$ are updated during the course of execution.

A conditional statement such as {\tt if} inside the target program may contain unknown values in its conditions.
Unknown branches during the execution interrupt the execution since the executor does not know which direction to explore.
Instead of always following one path, which increases false negatives significantly, LeakSemantic adopts a depth-first search scheme while taking the symbolic path constraints of unknown variables into account to reduce the search space.

More specifically, whenever an unknown branch is encountered, LeakSemantic creates a snapshot to store the state of the executor and pushes the snapshot onto a stack \textit{SnapStack}.
The snapshot consists of a copy of 
the current running context including the program counter and the values in the stack and the heap, which enables the executor to restore the environment after the unknown branch is processed and continue the analysis where it was left off.
The executor then explores each direction under the branch one by one, while using \textit{SnapStack} to save and restore the environment. 

Consider again the code shown in Listing~\ref{lst:dfs}.
The execution starts with an empty symbolic state and a symbolic path constraint $true$.
As a result, $\sigma = {mRun \mapsto mRun_0}$, where $mRun_0$ is an initially unconstrained symbolic value.
At every unknown conditional statement \textit{if} ($e$) \textit{then} S1 \textit{else} S2, $PC$ is updated to $PC \wedge \sigma(e)$ for the \textit{then} branch and $PC \wedge \neg \sigma(e)$ for the \textit{else} branch.
For instance, at the unknown condition in line 3, a snapshot of the executor is saved.
The executor first updates the $PC$ to $mRun_0 \not= null$ and explores the \textit{else} branch of the condition.
Once the execution terminates, it restores the status from the snapshot and proceeds to the \textit{then} branch of the condition in line 3 with $PC$ updated to $mRun_0 = null$.
The branch consists of a method \texttt{reportState()} that stealthily exposes user's private data, and another unknown condition (line 5).
The procedure to handle the second unknown condition is similar to the first one.
In this case, however, the \textit{then} branch has the path constraint $mRun_0 = null \wedge mRun_0 \not= null$ leading to an infeasible path.
Therefore, the executor ignores the \textit{then} branch and only explores the \textit{else} branch.

\ignore{
\begin{figure}
\centering
\includegraphics[height=1.6in,width=0.4\textwidth]{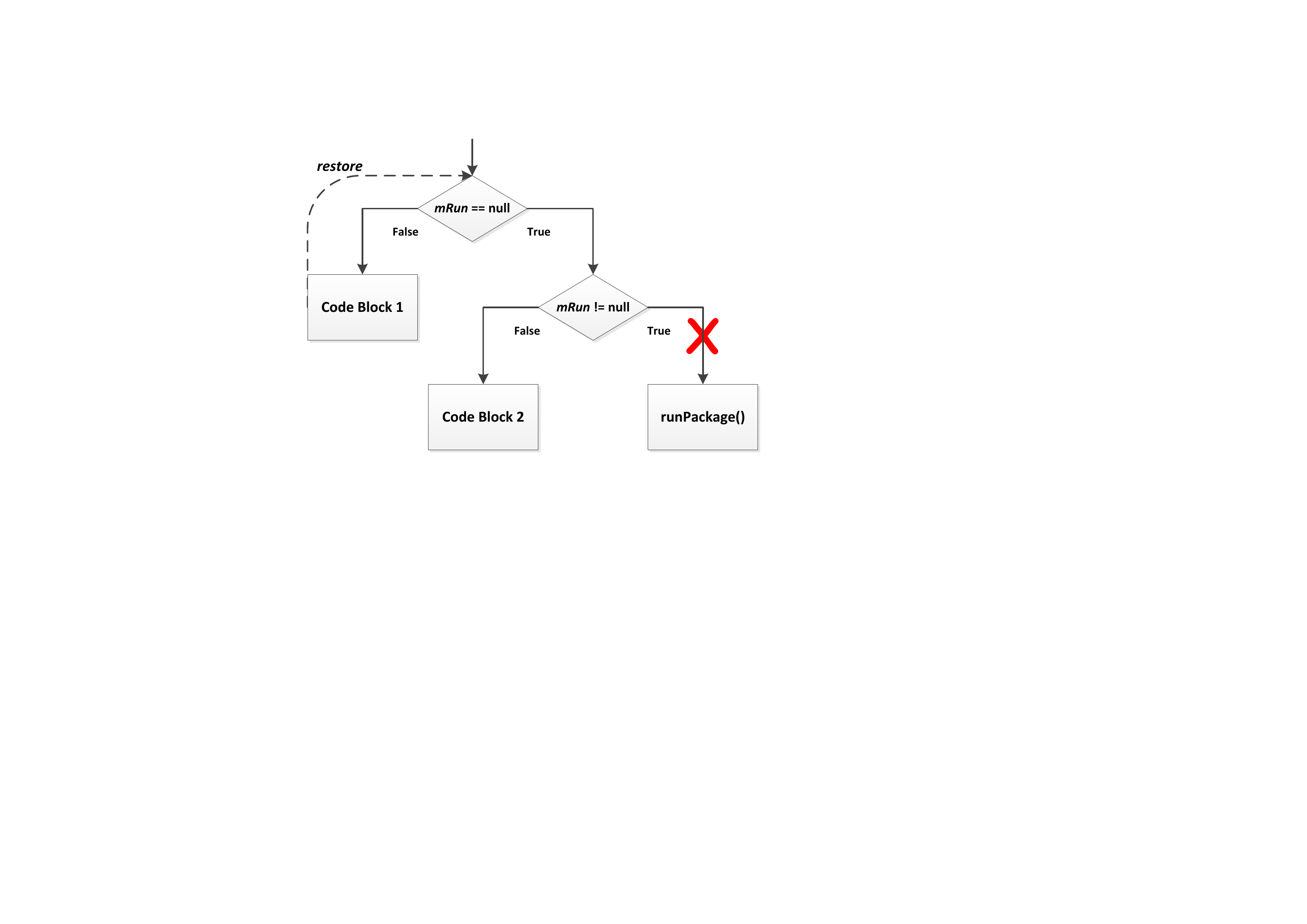}
\caption{Depth-first search}
\label{fig_dfs}
\end{figure}
}

Code containing loops or recursion may result in an infinite number of paths to be explored if the termination condition for the loop or recursion is symbolic.
Consider the code snippet shown below:
\begin{lstlisting}
        String[] x = getHttpResponse();
        int i = 0;
        while (!x[i].equals("")) {
          i++;
        }

        if (i > 3 && i < 10) {
          transmit(longitude, latitude);
        }
\end{lstlisting}
Since we do not know exactly how the server will respond in line 1, the content and the length of string array {\tt x} should be treated as unknown, leading to an infinite number of code paths.
To address this problem, 
previous studies~\cite{appaudit, rasthofer2016harvesting} simply set thresholds on analysis time or the number of visited instructions.
However, these approaches may lead to an incorrect value of {\tt i} after the loop, which should be equal to the actual length of {\tt x}.
Importantly, the value of {\tt i} is used to determine whether to trigger the leakage in line 7.

Instead, we execute the block under the loop only once and mark all the variables that accept new values within the block.
After exploration of the block, the tagged variables will be modeled symbolically for the rest of the execution.
By treating {\tt i} as a symbolic {\tt Integer} with constraint $i > 3 \wedge i < 10$, the sensitive transmission in line 8 will be successfully reached.
We also introduce some heuristics to further mitigate the issue of path explosion, which will be discussed in Section~\ref{sec:implementation}.

\subsection{Transmission Classification}\label{sec:classification}
Using the traffic flows generated by the dynamic analysis component, we formulate the detection of abnormal sensitive transmissions as a classification problem.
LeakSemantic uses a supervised learning approach to train classifiers that can be used by host-based or network-based intrusion detection systems. 
Specifically, we focus on \textbf{lexical features} derived from the set of URLs in the traffic traces.
Lexical features often contain useful patterns to distinguish between suspicious and benign traces.
URLs such as \texttt{gad.ju6666.com/GetAd?\&lo=(.*)} and \texttt{api.openweathermap.org/forecast?\&lon=(.*)}, in which {\tt lo} or {\tt lon} is an abbreviation of ``longitude", have the user's location data embedded.
The words {\tt GetAd} and {\tt forecast} further provide hints about the purposes of the transmissions:
the former URL is sent as a request for advertisement while the latter is composed to retrieve the corresponding weather forecast.
An effective detector should be able to report the ad request as suspicious and release the operational weather trace.

We utilize the simple yet powerful ``bag-of-words'' model~\cite{url-lexical} that is frequently used in spam detection to derive features inside URLs.
LeakSemantic divides a URL into tokens by treating certain characters as separators.
Each distinct token is then viewed as a separate feature and every data flow collected is then converted to a vector of binary values.
Direct application of ``bag-of-words" may produce a very large feature space, which results in a heavy computational cost.
As stated in~\cite{ren2016recon}, one can limit the size of the feature set by removing tokens that seldom appear in the flows.

\section{Implementation} \label{sec:implementation}
In this section, we provide further details about the implementation of LeakSemantic. 
LeakSemantic is mostly written in Java and consists of around 18,600 source lines of code.

LeakSemantic extends a part of FlowDroid for call graph generation.
\ignore{
Moreover, inter-component communication in Android is implemented through the use of \texttt{Intents}, with \texttt{IntentFilters} specifying the corresponding receivers.
To support inter-component communication, our system also parses the \texttt{Manifest.xml} file in the target application package to extract the declarations of \texttt{IntentFilters}. \red{not very clear how this is related to static or dynamic analysis.}
}
We implemented our own executor with taint analysis support to perform the dynamic analysis mentioned in Section~\ref{sec:dynamic}.
The executor leverages {\tt PATDroid\footnote{https://github.com/mingyuan-xia/PATDroid}} to extract bytecode and then interprets each bytecode instruction one by one.
During the execution, the sensitive data propagation is tracked by the taint analysis plugin.
Android applications invoke the APIs provided by the Android SDK to interact with the underlying operating system during runtime.
However, the official Android SDK is missing critical parts of the Android runtime, which are filled with ``stubs'' used for compilation.
The execution and taint analysis cannot proceed correctly without precisely modeling of the Android runtime.
We therefore manually pad the incomplete Android SDK and emulate the core functionalities offered by Android.
Our simulation of the Android system is similar to the Android Device Implementation (ADI) used in DroidSafe~\cite{gordon2015information}.
But their implementation is purely for static analysis and does not scale well to support our dynamic analysis.

LeakSemantic is currently using the \texttt{JaCoP}\footnote{https://jacop.osolpro.com/} to represent and update the path constraints.
To alleviate the path explosion caused by unknown branches, we heuristically limit the number of unknown variables.
We use the API \texttt{android.net.NetworkInfo/isConnected()} to illustrate the idea.
\texttt{isConnected()} reveals the real time connection capability of the device, so that the return value reflects the device status.
This should be treated as unknown in theory as mentioned in Section~\ref{sec:dynamic}.
However, the transmission can be triggered only if the device is connected to the Internet.
We therefore force the API call to always return \textit{true} instead.

We also simulate some commonly used third-party libraries to reduce performance overhead.
For instance, \texttt{com.squareup.picasso} is a widely used open-source package to support downloading and presenting images.
Since no misbehavior in it has been detected, we do not check the subroutines called by the package during execution.
Instead, we replace methods inside the official packages with our own methods during the execution.

\ignore{summary func: getStringByURL()}

\section{Evaluation} \label{sec:eval}
We have conducted a comprehensive evaluation of LeakSemantic. In this section, we report the evaluation results and our findings. Our evaluation contains two steps.
First, we leverage micro-benchmark suites to evaluate the leakage detection accuracy of our hybrid program analysis module.
Second, we apply LeakSemantic to real-world apps and construct classifiers to detect illegitimate exposures for different settings.
\ignore{The reachability of each connection point such as {\tt HttpClient.openConnection()} during the execution is a transmission. \red{hard to understand.}
A transmission that carries sensitive data is treated as a sensitive transmission.}
\vspace{-1.5ex}
\subsection{Benchmark Suites and Quality of Program Analysis}
We compared LeakSemantic with the following state-of-the-art taint analysis tools:
\begin{itemize}
  \item Andrubis~\cite{lindorfer2014andrubis} is a dynamic analysis sandbox based on TaintDroid.
  It generates nearly 8,000 pseudo-random streams of external events and monitors the behavior of the target app for 240 seconds\footnote{The official Andrubis service is no longer available. We installed TaintDroid on a real device and composed scripts to create an environment similar to Andrubis.}.
  \item FlowDroid is a flow-, field-, and object sensitive static program analysis framework.
  The original FlowDroid 
  cannot track information flows across separate components.
  We integrated FlowDroid with Epicc~\cite{li2015iccta} to partially support inter-component communications. 
  \item AppAudit is a hybrid taint analysis approach similar to LeakSemantic.
  It also uses static analysis to mark potential leaking methods, and then prune candidate methods through dynamic analysis.
  But the way it generates call graphs and models the unknown variables is different from LeakSemantic. 
\end{itemize}

We executed LeakSemantic on a computer with an Intel Core CPU E8500 @ 3.16GHz and 2GB of heap memory for the JVM.
Since Andrubis has fixed analysis time and AppAudit does not provide installation package to run locally, it is hard to compare the running times of the set of tools directly.
However, we observe that LeakSemantic exhibits good performance on the apps with short analysis time. 

We evaluated the detection accuracy of the above tools using the following two micro-benchmark sets.
LeakSemantic spent 12.4s on average for each app and FlowDroid took an average of 13.2s per app:
\subsubsection{DroidBench}
DroidBench\footnote{The up-to-date stable release is DroidBench 2.0 (https://github.com/secure-software-engineering/DroidBench/tree/master). We replaced all the sinks with network transmissions since Andrubis and AppAudit do not treat certain sinks as sensitive in some apps.} is an open-source benchmark suite that contains a set of hand-crafted apps that exploit various characteristics of the programming language to bypass static taint analysis.
It contains 118 apps in total, among which we excluded 10 apps with leakage types unsupported by both Andrubis and AppAudit, such as leaking user input passwords.

Table~\ref{tab:droidbench} summarizes the detection results over DroidBench. 
We observe that LeakSemantic achieves the best quality among the four taint analysis tools.
Precise call graphs and the better handling of unknowns enable LeakSemantic to generate zero false alarms.
Among the three baselines, Andrubis performs best and successfully report most leakages.
This is because the dataset is originally designed to test static analysis tools and difficulties for static analysis are typically not hard for dynamic analysis.
FlowDroid is able to locate more than 75\% of leaks. But its over-approximation also leads to the worst precision.
Also, FlowDroid is unable to generate runtime data such as traffic flow, and therefore cannot be directly used to build a traffic-based transmission classification model. 

Since both AppAudit and LeakSemantic adopt hybrid program analysis, we conducted a more detailed comparison between them.
LeakSemantic achieves better detection accuracy for several reasons.
First, AppAudit terminates its execution once a sink is touched.
As we discussed in Section~\ref{sec:dynamic}, reachability alone does not necessarily imply a sensitive transmission. 
Second, AppAudit does not consider some types of unknowns and always exploits one direction of an unknown branch, which introduces false negatives.
Moreover, LeakSemantic provides a more complete implementation of dynamic analysis to support various mechanisms used in Android.
In particular, LeakSemantic is able to locate event handlers registered in the layout configurations and track the communications among multiple components.
AppAudit does not support any of these Android features.
Last, as we mentioned in Section~\ref{sec:static}, the inaccurate model of call graphs used by AppAudit increases its false positives.

LeakSemantic (and all the three baselines) misses two flows that involve inter-application communications, which requires modeling the behaviors across multiple apps.
None of the existing taint analysis tools can detect this kind of collusion attack.
Another unresolved challenge of LeakSemantic is \emph{control flow dependent taints}, also a well acknowledged drawback in most taint analysis tools~\cite{appaudit}.
\ignore{
\footnote{Given TP (the number of true positives), FN (the number of false negatives), FP (the number of false positives) and TN (the number of true negatives), the efficiency of prediction of the model is measured based on the following metrics:\newline
TP Rate $= \frac{TP}{TP+FN}$, \quad FP Rate $= \frac{FP}{FP+TN}$, \newline
Precision $= \frac{TP}{TP+FP}$, \ \ F-measure $= \frac{2TP}{2TP+FP+FN}$.}
}

\begin{table}[t]
\centering
\begin{threeparttable}
\caption{\label{tab:droidbench}Detection results on DroidBench}
\small{
\begin{tabular}{ l  c  c  c  r }
\toprule
Tools & Missed Flows & Accuracy & FP & Precision \\\midrule
Andrubis & 15 & 84.2\% & 0 & 100\% \\
FlowDroid & 22 & 76.8\% & 10 & 56.6\% \\
AppAudit & 56 & 41.1\% & 2 & 91.3\% \\
LeakSemantic & 2 & 97.9\% & 0 & 100\% \\
\bottomrule
\end{tabular}
}
\end{threeparttable}
\begin{tablenotes}
\item \footnotesize FP = False Positives
\end{tablenotes}
\end{table}

\subsubsection{BombBench}
BombBench\footnote{https://github.com/bombbench/BombBench} is another open-source benchmark that contains 22 apps to test taint analysis tools. 
Each app takes advantage of a kind of \emph{logic} or \emph{time} bomb inspired by previous studies~\cite{zhang2014semantics, yangappcontext, rasthofer2016harvesting}
to conceal a sensitive flow.
We show the results in Table~\ref{tab:bombbench}.
LeakSemantic identifies most leaks among all the four tools.
We can see the sharp decrease of accuracy in Andrubis, which indicates that current random-events based testing toolkit is not powerful enough to cover complicated program logic.
Its limitation is fundamental and cannot be simply settled with extension of analysis time.
For example, \texttt{DevInfo2} triggers its payload only under certain system language.
Because, unlike LeakSemantic, they do not count as unknown the variables obtaining values from {\tt Locale/getDisplayName()},
both Andrubis and AppAudit fail to capture the disclosure flow.
We notice that FlowDroid also could not successfully mark this case, which may be caused by inaccurate modeling of system functions.
LeakSemantic missed one flow because of a variable implicitly assigned by a user-driven event.
Although we model variables who read the values from the UI-related API calls such as {\tt EditText/getText()} as unknowns,
currently we do not directly view the variables modified by the callbacks such as \texttt{onClick()} as unknowns even they are correlated with user interactions.
We do this for performance concerns since there might be plenty of variables influenced by the callbacks in real apps.
Excessive amount of unknowns leads to the exponential size of code paths needed to be explored.

\begin{table}[t]
\caption{Accuracy on BombBench}
\centering
\small{
\begin{tabular}{ l  c  r }
\toprule
Tools & Missed Flows & Accuracy \\\midrule
Andrubis & 21 & 4.5\% \\
FlowDroid & 14 & 36.4\% \\
AppAudit & 12 & 45.5\% \\
LeakSemantic & 1 & 95.5\% \\
\bottomrule
\end{tabular}
}
\label{tab:bombbench}
\end{table}

\subsection{Real Apps and Transmission Classification}
We then applied LeakSemantic to build a traffic classification model using real apps.
From the traffic generated by our hybrid analysis tool, it is possible that multiple code paths lead to the same connection, which results in separate transmissions with an identical URL. We merged these transmissions with the same URL into a single one within the target app.

\begin{figure}
\centering
\includegraphics[height=1.55in,width=0.36\textwidth]{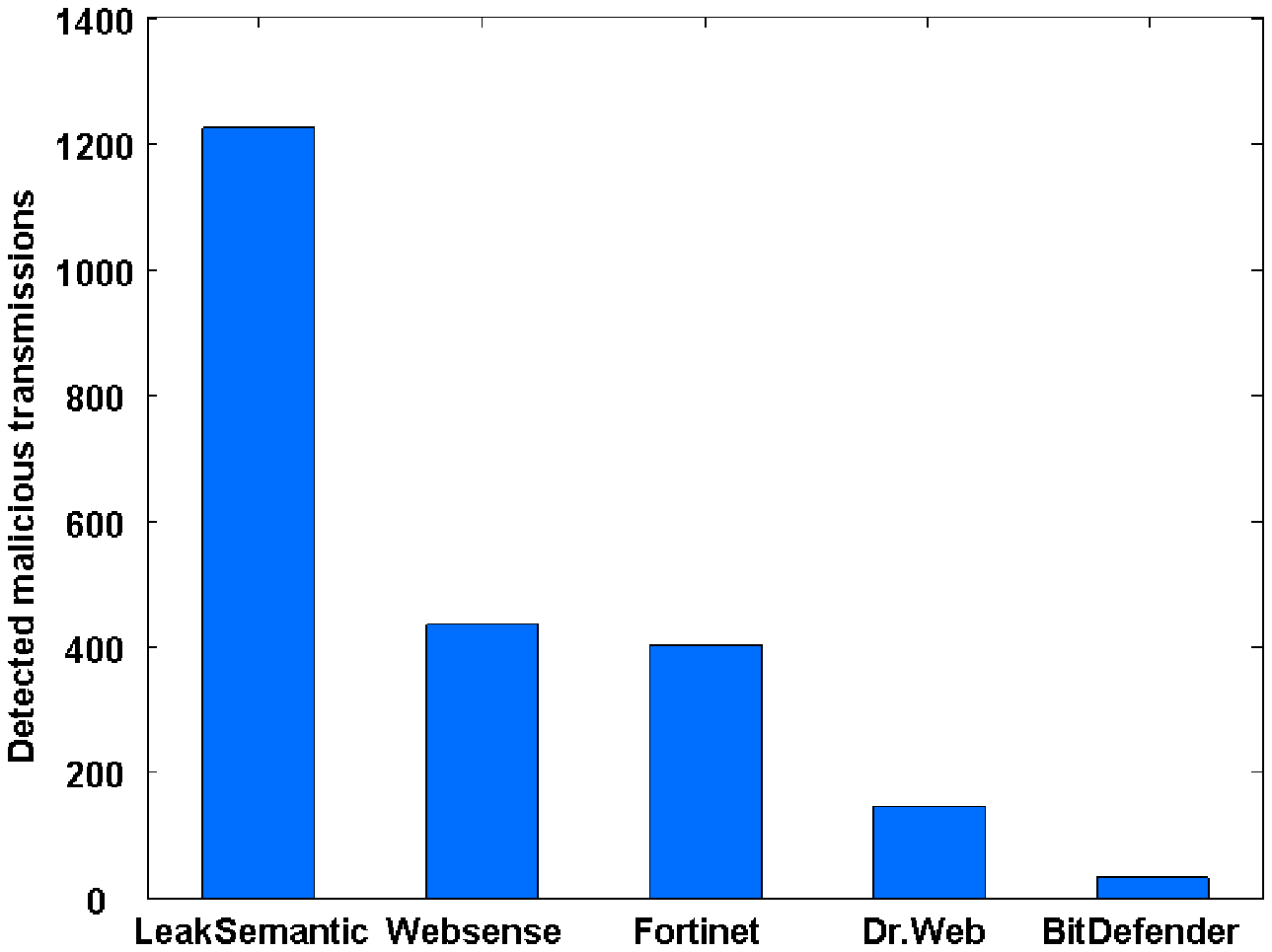}
\vskip-1em
\caption{\small Detected malicious sensitive transmissions.}
\label{fig_virustotal}
\end{figure}

We first collected malicious sensitive transmission from the Android Malware Genome project, which contains 744 leaking malwares~\cite{zhou2012dissecting}.
LeakSemantic extracted 1223 malicious sensitive transmissions and collected the corresponding traffic. 
We first observe that these malicious transmissions cannot be correctly identified by existing commercial anti-virus solutions, which motivates the need for a new detection approach. 
To this end, we uploaded the URLs of these transmissions to \texttt{VirusTotal}\footnote{https://www.virustotal.com/}, a popular website that scans submitted URLs with latest 68 anti-virus engines.
Surprisingly, 64 out of 68 engines did not report any alarms regarding the transmissions.
Figure~\ref{fig_virustotal} presents the detection results by the rest 4 engines.
\texttt{Websence} identified relatively more malicious URLs (436, or 35.7\%), 
but the number found is still far from 50\% of all malicious connections.

We then ran LeakSemantic on 660 apps crawled from the categories that have legal sharing functionalities in app markets\footnote{Google Play (https://play.google.com/store/apps) and Baidu App Market (http://shouji.baidu.com)}.
Among them, LeakSemantic recognized 1056 sensitive transmissions. The average analysis time for each app is 135.3s, including the 744 malwares and the 660 authentic apps.
For each flow collected, we examined the destination host name.
If the host name belongs to an advertisement or analytics server, we marked the flow as illegal.
We then checked the plain text content delivered through the flow to see whether the response sent by the server is related to the sent user data or not.
There are cases in which the communication between the phone and the server are encrypted.
We leveraged instrumentation and reverse engineering to block those flows.
We reran the modified app to see how blocking influences the app.
The flow was labeled as legal when the app's functionality is affected.
Out of 1056 transmissions, 791 did not affect the app's functionality, so we labeled them as illegitimate.
The other 265 operational sensitive transmissions were collected from 183 apps.

We used the labeled 2279 transmissions as training and testing data with ten-fold cross-validation~\cite{kohavi1995study}, which is a standard approach for evaluating machine learning solutions.
We applied \textbf{Decision Tree} as the learning classifier for LeakSemantic since it is commonly used in traffic classification~\cite{ren2016recon, raghuramu2015uncovering}.
As mentioned earlier in Section~\ref{sec:intro}, LeakSemantic can be deployed as a host-based or network-based detection system.
We conducted two experiments that reflected the effectiveness of LeakSemanic in different scenarios.
When LeakSemantic is configured in a single host system, it automatically finds the disclosure points and then picks the illegal instances based on the flows generated.
The classifier at host-level involves only the flows of sensitive transmissions;
the detection model at network-level should be able to filter out the innocent flows that do not carry any sensitive data.

\subsubsection{Host-based Detection}
Table~\ref{tab:host} shows that LeakSemantic has high precision and F-measure in identifying illegal transmissions\footnote{Since the data is heavily skewed towards the illegal class, we used SMOTE~\cite{chawla2002smote} to over-sample the legitimate class.}.
After manually inspecting the misidentified instances, we found that their URLs were very similar to the benign addresses.
Also, they put the sensitive data into their body rather than the URL, which makes the URL-based detection more difficult to correctly label them.
We note that LeakSemantic is able to collect more information than URLs. We plan to consider more features to further reduce the false negatives in the future.

\begin{table}[t]
\centering
\caption{Host-based Classification Results}
\label{tab:host}
\small{
\begin{tabular}{ c  c  c  c  c }
\toprule
Class & TP Rate & FP Rate & Precision & F-measure\\\midrule
Illegal & 0.938 & 0.063 & 0.974 & 0.956 \\
Legal & 0.937 & 0.062 & 0.856 & 0.895 \\
\bottomrule
\end{tabular}
}
\end{table}

\subsubsection{Network-based Detection}
Based on the sensitive transmissions we collected, we added the non-sensitive traffic flows to the legitimate class.
This reflects the real environment of the network-based detection.
Table~\ref{tab:network} summarizes our results.
As we can see, the prediction incurs a slight loss in accuracy compared to the results of the host-based detection.
This is expected as the addition of non-sensitive flows makes the learning task more challenging. 

\begin{table}[t]
\centering
\caption{Network-based Classification Results}
\label{tab:network}
\small{
\begin{tabular}{ c  c  c  c  c }
\toprule
Class & TP Rate & FP Rate & Precision & F-measure\\\midrule
Illegal & 0.915 & 0.095 & 0.916 & 0.915  \\
Legal & 0.905 & 0.085 & 0.904 & 0.904  \\
\bottomrule
\end{tabular}
}
\begin{tablenotes}
\item \footnotesize TP = True Positive, FP = False Positive
\end{tablenotes}
\end{table}

During the experiments, we also observed the following interesting phenomena:

\noindent \textbf{Finding 1:}
Among the 1223 malicious leaking transmissions extracted from the malware dataset, we found that 69.7\% of the transmissions used encryption to hide the hostnames.
Malware leverages encryption to evade traditional signature-based detection approaches. 
As mentioned earlier, encryption also hinders pure static analysis from explicitly detecting the target behaviors.
Without enough dynamic information, the intrusion detection systems failed to locate many malicious transmissions.
To illustrate how important the decryption is, we conducted an experiment that trained a model based solely on unencrypted instances and tested the model on the instances with encrypted hostnames.
Among the 806 encrypted instances, the model only recognized 578 (71.7\%) of them.
Compared to the prediction results (91\%) shown previously, the accuracy decreased dramatically.

\noindent \textbf{Finding 2:}
LeakSemantic identified more than 1223 malicious transmissions in the malware dataset.
However, it could not properly generate traffic flows for a few transmissions such as those from the \texttt{DroidKunfu4} malware family.
We manually inspected the code and found that the hostnames of the transmissions are not embedded either in the code or in the resource files of the apps.
Instead, the transmissions dynamically retrieve the hostnames from a remote server with the help of the command and control modules.

\noindent \textbf{Finding 3:}
From the crawled apps, we noticed that 3 connections indirectly leak the private data.
Instead of sending the user data directly to a tracing server, they first grab the user's coordinates and query a legitimate popular location server to get the corresponding description.
They then transmit the description to a suspicious server.
Such behavior suggests the need to track the influence of a connection even when the first connection contacts a legitimate server.

\noindent \textbf{Finding 4:}
LeakSemantic found no sensitive HTTPS connections in the malwares.
However, 27 illegitimate HTTPS transmissions were identified in the authentic apps and they were all built by third-party ads/analytics libraries.
Although sensitive HTTPS connections are not popular at the current time, we foresee the necessity of inspecting HTTPS connections with the techniques such as \texttt{SSLsplit}\footnote{https://www.roe.ch/SSLsplit} in the future.
\ignore{eavesdropping}

\ignore{loc vs imei number.}

\noindent \textbf{Finding 5:}
We found that more than 60\% of the 183 apps that have legitimate sharing connections also contain illegal transmissions inside for ad or analytics purposes.
We also found a weather application that only transmits users' location data to ad servers.
It is highly probable that the users of these apps will grant the app the permission to access sensitive resources without knowing their private data will be collected stealthily by unintended servers.

\vspace{-0.45ex}
\section{Limitations}\label{sec:limitation}
Our approach has the following limitations:
\begin{itemize}
\item If an adversary knows our approach, he could obfuscate the flows to match our criteria.
We envision that more features need to be considered in the future. 
\item The technique most closely related to our dynamic analysis is \emph{concolic testing}~\cite{godefroid2007compositional}, which also leverages both concrete and symbolic values to proceed its execution.
Our approach inherits its path explosion limitation; the size of code paths is exponential in the number of unknown branches.
We currently remove most unnecessary unknowns with our specific preprocessing and we will look into more advanced relevant techniques soon.
\ignore{\item Although it is rare, an adversary can hide the dynamic information?}
\end{itemize}

\section{Related Work}\label{sec:related}
Dynamic and static taint analysis track sensitive data flows in programs.
TaintDroid~\cite{enck2014taintdroid} modifies the Dalvik virtual machine to monitor potential leaks at runtime.
It only identifies leakage that is actually triggered during execution, thus requiring a driver with good code coverage.
The static analysis tools FlowDroid~\cite{arzt2014flowdroid} and DroidSafe~\cite{gordon2015information} overcome the coarse-granularity through over-approximation.
But they also suffer from imprecision by visiting code paths that are not actually feasible.
AppAudit~\cite{appaudit} leverages hybrid static-dynamic analysis in order to keep the advantages and avoid the drawbacks of both. 
It only examines code paths determined statically and explores one path when it encounters an unknown branch.
In contrast, our system dynamically extends the code coverage and explores as many paths as feasible when an unknown branch is found.
ReCon~\cite{ren2016recon} is a solely network-based detection that learns patterns from traffic traces, which is similar to the transmission classification used in LeakSemantic.
Our program analysis approaches can further improve the performance of network-based detection.
All above approaches treat any exposures of user data as illegitimate, which obscure the true threats through generating large number of false alarms.

AppIntent~\cite{yang2013appintent} first stresses the necessity to justify the sensitive transmissions in apps.
Bayesdroid~\cite{tripp2014bayesian} proposes a solution by treating the transmissions that carry less accurate information as legal.
However, a transmission could be very harmful even if it only contains coarse information since it can collude with others.
FlowIntent~\cite{fu2016flowintent} leverages front-page user interfaces to discriminate location-sharing communications.
Its effectiveness depends on the content shown on the pages and its underlying random fuzzing based approach, which is similar to Andrubis~\cite{lindorfer2014andrubis}, makes it hard to locate stealthy malicious payloads.
AAPL~\cite{lu2015checking} is a static app auditing tool that queries a commercial recommendation system to rank sensitive disclosures.
But as shown in~\cite{fu2016flowintent}, being in the same category does not imply having the same functionality. 
Other static analysis approaches including AsDroid~\cite{huang2014asdroid} and DroidJust~\cite{chen2015droidjust} only treat connections that do not influence the user-observable phone states as malicious.
But a flow can still be malicious even it leads to visible changes as it can also trigger the underlying malicious payload simultaneously.
LeakSemantic looks beyond the mere surface of leaks by examining their intention based on the corresponding traffic flows.

\section{Conclusion}\label{sec:conclusion}
In this work, we developed a prototype called LeakSemantic that can identify suspicious sensitive network transmissions from mobile apps automatically.
Its hybrid program analysis component enables it to provide better accuracy and precision than other state-of-the-art taint analysis approaches.
LeakSemantic further constructs machine learning classifiers to differentiate among the disclosures based on features derived from the program analysis.
Our evaluation on 2279 sensitive connections collected from real-world 1404 apps shows that LeakSemantic achieves a detection accuracy of 91\%.

\section{ACKNOWLEDGEMENTS}
The effort described in this article was partially sponsored by the U.S. Army Research Laboratory Cyber Security Collaborative Research Alliance under Contract Number W911NF-13-2-0045.
The views and conclusions contained in this document are those of the authors, and should not be interpreted as representing the official policies, either expressed or implied, of the Army Research Laboratory or the U.S. Government.
The U.S. Government is authorized to reproduce and distribute reprints for Government purposes, notwithstanding any copyright notation hereon.


\footnotesize
\bibliographystyle{abbrv}
\bibliography{ref}

\end{document}